\documentstyle[pra,aps,multicol]{revtex}

\newcommand{\be}{\begin{equation}}
\newcommand{\ee}{\end{equation}}
\newcommand{\barx}{\bar{x}}
\newcommand{\ttheta}{\tilde{\theta}}
\newcommand{\tOmega}{\tilde{\Omega}}

\begin{document}
\draft
\title{Periodic Hamiltonian and Berry's phase in harmonic oscillators}
\author{Dae-Yup Song}
\address{ Department of Physics,\\ Sunchon National University, Sunchon 
540-742, Korea}
\date{\today}
\maketitle
\begin{abstract}
For a time-dependent $\tau$-periodic harmonic oscillator of two linearly
independent homogeneous solutions of classical equation of motion which are
bounded all over the time (stable), it is shown, 
there is a representation of states cyclic 
up to multiplicative constants under $\tau$-evolution or $2\tau$-evolution 
depending on the model. The set of the wave functions is complete. 
Berry's phase which could depend on the choice of representation can be 
defined under the $\tau$- or $2\tau$-evolution in this representation.
If a homogeneous solution diverges as the time goes to infinity, it is 
shown that, Berry's phase can not be defined in any representation considered.
Berry's phase for the driven harmonic oscillator is also considered.
For the cases where Berry's phase can be defined, the phase is given 
in terms of solutions of the classical equation of motion.

\end{abstract}

% body of paper here
\begin{multicols}{2}
\section{Introduction}

Berry \cite{Berry,SW} showed that the cyclic adiabatic change of Hamiltonian 
induces, in the phase of wave function, a change which separates into the 
obvious dynamical part and an additional geometric part.
Aharonov and Anandan \cite{AA} generalized Berry's result to 
nonadiabatic cases, by giving up the assumption of adiabaticity. 
The price we have to pay in making this generalization is that quantum 
states after the evolution of Hamiltonian's cycle are not necessarily 
equivalent to the original states up to phase, so that Berry's phase could 
in general not be defined. A specific example of this failure in (quasi)cyclic 
evolution of quantum states is given in a driven harmonic oscillator 
system where the particular solution diverges as time goes 
to infinity \cite{Moore}.

The (driven) harmonic oscillator system of time-dependent mass and/or 
frequency has known to be a system whose quantum states
can be given in terms of the solutions of classical equation of motion
\cite{Lewis,Song1,KL,Ji}.  
The time-dependent $\tau$-periodic (driven) harmonic oscillator system  
is considered in Refs.\cite{Moore,Ji} based on the 
Floquet theory; If the solutions of classical equation of motion are 
stable (finite all over the time), it has been shown 
{\it with other assumptions} that \cite{Ji}, there might exist positive integer 
$N$ such that wave functions are quasiperiodic under the $N\tau$-evolutions. 
In those cases, the Berry's phases were given in terms of auxiliary functions related to 
the solutions of classical equation of motion. 

The driven harmonic oscillator is described by the Hamiltonian:
\be
H^F(x,p,t)= {p^2 \over 2 M(t)} +{1\over 2} M(t)w^2(t)x^2-xF(t),
\ee
with positive mass $M(t)$, real frequency $w(t)$ and external force $F(t)$. 
For these parameters, we require a periodicity so that
\be
H^F(x,p,t+\tau)=H^F(x,p,t).
\ee
The classical equation of motion of the system is given as 
\be
{d \over {dt}} (M \dot{\barx}) + M(t) w^2(t) \barx=F(t).
\ee
General solution of Eq.(3) is described by two linearly independent homogeneous
solutions $\{u(t),v(t)\}$ and a particular solution $x_p(t)$. 
For convenience, we define $\rho(t)$ and time-constant $\Omega$ as
\begin{eqnarray}
\rho(t)&=&\sqrt{u^2(t) +v^2 (t)},\\
\Omega&=&M(u \dot{v} -v\dot{u}).
\end{eqnarray}
It is known that \cite{Song1,Song2} the wave functions satisfying the 
corresponding Schr\"{o}dinger equation are given as
\begin{eqnarray}
\psi_n^F &=&\cr
    && {1\over \sqrt{2^n n!}}({\Omega \over \pi\hbar})^{1\over 4}
     {1\over \sqrt{\rho(t)}}[{u(t)-iv(t) \over \rho(t)}]^{n+{1\over 2}}\cr
  &&\times   \exp[{i\over \hbar}(M\dot{x}_px+\delta(t))] 
\cr&&
       \times \exp[{(x-x_p)^2\over 2\hbar}(-{\Omega \over \rho^2(t)}
               +i M(t){\dot{\rho}(t) \over \rho(t)})]
\cr&&  
       \times    H_n(\sqrt{\Omega \over \hbar} {x -x_p \over \rho(t)}), 
\end{eqnarray}
where $\delta(t)$ is defined as 
\be 
\delta(t)=\int_{0}^t[{M(z)w^2(z)\over 2} x_p^2(z) 
                -{M(z) \over 2}\dot{x}_p^2(z) ] dz.
\ee
Different choice of $\{u,v\}$ gives different set of wave function, while 
a set with two linearly independent $\{u,v\}$ is complete in the sense that
\begin{eqnarray}
K(x_b,t_b;x_a,t_a)&=&\sum_n \psi_n^F(x_b,t_b){\psi_n^F}^*(x_a,t_a)\cr
&\rightarrow& \delta(x_b-x_a) 
~~~{\rm as}~t_b\downarrow t_a.
\end{eqnarray}
From the wave functions in Eq.(6), one can easily find that, the wave functions
become (quasi)periodic when the $\rho(t)$ and $x_p(t)$ are periodic with a 
common period.

In this paper, we will show that, if there exist $u(t),v(t)$ which are 
stable, for the oscillator without driving force 
there is a representation of (quasi)periodic wave functions
under $\tau$-evolution or $2\tau$-evolution depending on the model. 
If one of the homogeneous solution diverges as the time goes to infinity, 
it will be shown that, Berry's phase can not be defined in any representation 
and the uncertainty of position diverges in the limit.
For the oscillator with driving force, it will be shown that, if 
there exist two homogeneous stable solutions and ${p \over N}\tau$-periodic  
particular solution with two positive integers $p$ and $N$
of no common divisor except 1, the wave functions are (quasi)periodic under 
$N\tau$-  or $2N\tau$-evolution so that Berry's phase can be defined.
For the cases where Berry's phase can be defined, the phase will be given 
in terms of solutions of the classical equation of motion.

In the next section, the harmonic oscillators without driving force 
will be considered. In section III, the driven harmonic oscillators will 
be considered. In section IV, it will be shown that if one of 
homogeneous solutions diverges, Berry's phase can not be defined 
in any of the representations. A summary will be given in the last section.

\section{The Hill's equation and quasiperiodic quantum states}
In this section we will consider the harmonic oscillator without 
driving force. The wave functions $\psi_n(x,t)$ of the case are given
from $\psi_n^F(x,t)$ in Eq.(6) by letting $x_p=\delta=0$.
\subsection{The Hill's equation}
We start with the case of unit mass. For this case, the classical equation 
of motion of Eq.(3) reduces to the Hill's equation:
\be
\ddot{\barx} +  w_0^2(t) \barx =0,
\ee
which has long been studied in mathematics \cite{Magnus}. 
The subscript 0 will be used to denote that the quantity
is defined for the system of unit mass, and thus the two linearly independent 
solutions of Eq.(9) will be denoted as $\{u_0,v_0\}$. 
In considering Eq.(9) as an equation of motion of a classical system, 
one may naively expect that the force varying periodically with 
time acts on the (unit) mass in such a manner that the force tends to move the
mass back into a position of equilibrium ($\barx=0$) in proportional to 
dislocation of the mass, so that the mass is confined to a neighborhood of 
$\barx=0$. Through the analyses of Hill's equation, however, it turns out 
that this expectation needs {\it not} to be always the case. 
In fact, an increase of the restoring force may cause the mass to oscillate
with wider and wider amplitude, as can be seen from the Liapounoff's theorem
\cite{Lia,Magnus}.

A detail information on whether the Hill's equation has an unstable 
solution can be obtained from the Floquet's theorem \cite{Magnus}.
If a constant $\alpha$ which is called {\em characteristic exponent}
is not one of the $m\pi/\tau$ 
$(m=0,\pm 1,\pm 2,\cdots)$, the theorem states that two linearly 
independent solutions of Hill's equation are written as 
\be
\barx_1(t)= e^{i\alpha t} p_1(t),~~~~\barx_2(t)= e^{-i\alpha t} p_2(t)
\ee
where $p_1(t)$, $p_2(t)$ are periodic functions with period $\tau$. 
In the case that $\alpha$ is one of the $m\pi/\tau$, the theorem states,
both of the two linearly independent solutions of Hill's equation are  
stable if and only if both of them can be written as periodic functions 
of period $\tau$ or $2\tau$ depending on $w^2_0(t)$.

If the two linearly independent solutions are written as in 
Eq.(10) and both of them are not $\tau$-periodic nor $2\tau$-periodic,  
they are stable if $\alpha$ is real \cite{Magnus}. In this stable case, 
by combining linearly $\barx_1(t)$ and $\barx_2^*(t)$ with complex 
coefficients and taking real, imaginary part of the new solution
as another set of solutions, one can always find two linearly 
independent real solutions $u_0(t),v_0(t)$ of Hill's equation as
\be
u_0(t)=Ap(t)\cos(\alpha t + \ttheta (t)),~
v_0(t)=Bp(t)\sin(\alpha t + \ttheta (t)),
\ee 
with $\tau$-periodic real functions 
$p(t),\ttheta(t)$ and nonzero constants 
$A,B,\Omega_0~ (\equiv ABp^2(\alpha+\dot{\ttheta}))$.

\subsection{Quasiperiodic quantum states}
It is known that \cite{Magnus,Song2}, if 
\be
w_0^2(t)=w^2 -{1\over \sqrt{M}}{d^2\sqrt{M}\over dt^2},
\ee
$\{u(t),v(t)\}$ is given from $\{u_0(t),v_0(t)\}$ as
\be
u(t)={u_0(t) \over \sqrt{M}},~v(t)={v_0(t) \over \sqrt{M}}.
\ee

If $\alpha$ is one of the $m\pi/\tau$ $(m=0,\pm 1,\pm 2,\cdots)$ and the 
two linearly independent solutions are stable, then  $\{u,v\}$ are $\tau$- 
or $2\tau$-periodic. In order to find the overall phase change, we need to 
consider the complex $z$-plane of $z=u+iv$. In the plane, the trajectory 
of $z(t)$ makes a closed curve $C$, since $z(t+\tau')=z(t)$ where $\tau'$ 
is $\tau$ or $2\tau$ depending on the periodicity of $\{u,v\}$. Making use 
of residue theorem, the number that the curve winds the origin can be shown to 
be equal to ${1\over 2\pi i}\oint_C {dz \over z}$. From this considerations 
one can find that
\be
\psi_n(x,t+\tau')
=e^{-i(n+{1\over 2})\int_0^{\tau'}{\Omega \over M\rho^2}dt}\psi_n(x,t).
\ee
Therefore the overall phase change can be written as 
\be
\chi_n=-(n+{1\over 2})\int_0^{\tau'}{\Omega \over M\rho^2}dt.
\ee

If $\alpha$ is not one of the $m\pi/\tau$ $(m=0,\pm 1,\pm 2,\cdots)$ 
and is real, by taking $A=B$, one can find the 
quasiperiodic wave functions under $\tau$-evolution. 
If $A=B$, the wave functions do not depend on the magnitude of $A$ (or $B$).
In the representation of $A=B$, the overall phase change under the $\tau$-evolution
is written as
\be
\chi_n= -(n+{1\over 2})\alpha \tau 
~~(=-(n+{1\over 2})\int_0^{\tau}{\Omega \over M\rho^2}dt).
\ee 

\subsection{Berry's phase}
The dynamical phase change $\delta_n$ during $\tau'$-evolution is given as
\begin{eqnarray}
&\delta_n(\tau')&\cr
&=&-{1\over \hbar}\int_0^{\tau'}\int_{-\infty}^\infty
      \psi_n^*(x,t) H(x,t) \psi_n(x,t) dx dt \cr
&=&-(n+{1\over 2})\int_0^{\tau'}
         [  {\Omega \over 2M(t)\rho^2(t)}
                           (1+{M^2(t) \over \Omega^2}\rho^2(t)\dot{\rho}^2 )\cr
&&~~~~~~~~    +{\rho^2(t)  \over 2\Omega}M(t)w^2(t)] dt. 
\end{eqnarray}

The Berry's phase $\gamma_n$ is given from the overall phase change by 
subtracting the dynamical change;
\be
\gamma_n=\chi_n - \delta_n.
\ee

If  $\alpha$ is one of $m\pi/\tau$ $(m=0,\pm 1,\pm 2,\cdots)$ and both of the 
linearly independent classical solutions are stable, the wave functions are 
$\tau$- or $2\tau$-periodic. In this case, the Berry's phase is given as 
\be
\gamma_n={1 \over 2} (n+{1\over 2})\int_0^{\tau'} 
      ({M \dot{\rho}^2 \over \Omega} - {\Omega\over M \rho^2}
         +{Mw^2 \over \Omega}\rho^2) dt,
\ee  
where $\tau'$ can be  $\tau$ or $2\tau$ depending on the model.

If  $\alpha$ is not one of $m\pi/\tau$ $(m=0,\pm 1,\pm 2,\cdots)$ and is real,  
the Berry's phase in the representation of $A=B$ is written  as 
\begin{eqnarray}
\gamma_n &=&-{1\over 2}(n+{1\over 2})\alpha\tau   \cr
&&~~~~+{1\over 2}(n+{1\over 2})\int_0^\tau [{w^2(t)p^2(t) \over \tOmega}
            +{M(t) \over \tOmega} \dot{q}^2]dt,
\end{eqnarray}
where 
\be
\tOmega={\Omega_0 \over A^2}~~{\rm and}~~q(t)={p(t)\over \sqrt{M(t)}}.
\ee 

For the $n=0$ cases, these results exactly agree with that of Ref.\cite{GC}.

A point that should be mentioned is that every phase is  
defined up to additive constant $2\pi$.

\section{Driven harmonic oscillator}

In this section, we will consider the driven harmonic oscillator.
Due to the lack of understanding on the periodicity of particular 
solution, our attention will be limited in the cases in which one can 
construct a particular solution periodic with period $r\tau$, where 
$r$ is written as $p/N$ with two positive integers $p$ and $N$
of no common divisor except 1. We will also restrict our attentions on the 
cases of the two linearly independent stable homogeneous solutions.

For a $\tau'$-evolution, the dynamical phase is written as
\begin{eqnarray}
&\delta_n^F(\tau')& = -{1\over \hbar}\int_0^{\tau'} <n|H^F| n> dt \cr
 &=& -{1 \over 2} (n+{1\over 2})\int_0^{\tau'} 
      ({M \dot{\rho}^2 \over \Omega} + {\Omega\over M \rho^2}
         +{Mw^2 \over \Omega}\rho^2) dt \cr
&&  -{1\over \hbar}\int_0^{\tau'} 
     ({3 \over 2} M\dot{x}_p^2 - {1 \over 2} Mw^2x_p^2) dt.
\end{eqnarray}

In the case that two linearly independent homogeneous solutions are
periodic under $\tau$- or $2\tau$-evolution by defining $\tau'$ as 
$N\tau$ or $2N\tau$ depending on the periodicity of the 
classical solutions, one can find the relation 
\be
\psi_n^F(t+\tau')= e^{-i(n+{1\over 2})\int_0^{\tau'}{\Omega \over M\rho^2}dt
         +i{\delta(\tau')\over \hbar}}\psi_n^F(t),
\ee 
which gives the overall phase change:
\be
\chi_n^F=-(n+{1\over 2})\int_0^{\tau'}{\Omega \over M\rho^2}dt
          +{\delta(\tau')\over\hbar}.
\ee
Berry's phase is thus given as
\begin{eqnarray}
\gamma_n^F(\tau') &=& {1 \over 2} (n+{1\over 2})\int_0^{\tau'} 
      ({M \dot{\rho}^2 \over \Omega} - {\Omega\over M \rho^2}
         +{Mw^2 \over \Omega}\rho^2)dt \cr   
  && + {1\over \hbar}\int_0^{\tau'} M\dot{x}_p^2 dt.
\end{eqnarray}

In the case that two linearly independent homogeneous stable solutions 
are not periodic under $\tau$- or $2\tau$-evolution, by letting $A=B$
we can find quasiperiodic wave functions under the $N\tau$-evolution with
the overall phase change:   
\be
\chi_n^F= -(n+{1\over 2}) \alpha N\tau 
        +{1\over \hbar}\delta(N\tau).
\ee 
In this representation, Berry's phase is given as 
\begin{eqnarray} 
\gamma_n^F &=& 
      -(n+{1\over 2}) \alpha N\tau \cr
&&   +{1 \over 2}(n+{1\over 2})N
      \int_0^{\tau}[{w^2 \over \tilde{\Omega}}p^2
                      +{M\dot{q}^2 \over\tilde{\Omega}}]dt   \cr
&&   +{1\over \hbar}\int_0^{N\tau} M\dot{x}_p^2 dt  
\end{eqnarray}
The last terms in the right hand side of Eqs.(25,27) are order of
$1/\hbar$; If they are zero, then $\gamma^F=N\gamma_n$.

\section{ Unstable classical solutions }
If the wave function is quasiperiodic, then $<n|x|n>$ and 
$<n|(\Delta x)^2|n>$  $(\equiv <n|x^2|n> - <n|x|n>^2)$ must be periodic.
For the driven case, if the $x_p$ diverges, 
because of the fact that $<n|x|n>=x_p$, the wave functions can not be 
quasiperiodic. 

For the driven or undriven case, $(\Delta x)^2$ is given as
\be 
(\Delta x)^2 = (n + {1\over 2}) {\hbar \rho^2 \over \Omega}.
\ee
If a homogeneous solution is unstable, then any set of two linearly 
independent homogeneous solution should have an unstable solution.
If one of the homogeneous solutions is unstable, Eqs.(4,28) show 
that $<n|(\Delta x)^2|n>$ can not be periodic and thus the wave functions 
can not be quasiperiodic. 

For the quasiperiodic wave functions, $<n|(\Delta p)^2|n>$ 
which is given as
\be
<n|(\Delta p)^2|n>= (n + {1\over 2})\hbar 
     [{\Omega\over \rho^2} + {M^2\dot{\rho}^2 \over \Omega}]
\ee
should also be periodic. In the case that imaginary part of $\alpha$
is not zero, one can easily show that  $<n|(\Delta p)^2|n>$ diverges 
in the limit $t\rightarrow \pm \infty$. In the case of unstable solution 
with $\alpha$ being one of the $m\pi/\tau$ $(m=0,\pm 1,\pm 2,\cdots)$,
$<n|(\Delta p)^2|n>$ remains finite in the limit, but does not converges
to 0 in general.

\section{Conclusions}

We have considered the (driven) harmonic oscillator system. 
For the $\tau$-periodic oscillator of two linearly 
independent homogeneous stable solutions without driving force, 
one can always construct a set of wave functions which are quasiperiodic 
under $\tau$- or  $2\tau$-evolution. 
The set of wave functions is complete.

If one of the homogeneous solution is unstable or particular 
solution of the driven system diverges, for the oscillator system with or 
without driving force, we prove that wave functions can {\it not} be 
quasiperiodic.

For the driven case, we illustrate the possibility of existence of
quasiperiodic wave functions under $N\tau$- or $2N\tau$-evolution which 
mainly depends on the behavior of the particular solution.
If the Berry's phase can be defined for a wave function of driven harmonic 
oscillator system, it must either be integral multiple of Berry's phase for 
the corresponding wave function without driving force or contain a term 
of $O(1/\hbar)$.

Recently Berry's phase of simple harmonic oscillator is studied in
\cite{PS} where it has been shown that the model can provide explicit 
examples for various cases considered here.

\acknowledgments
It is the author's pleasure to acknowledge helpful discussions with Prof. 
J.H. Park on mathematical aspects of the subject. This work is supported
in part by Non-Directed Research Fund, Sunchon National University.

\end{multicols}

\end{document}